\documentclass{aastex62} 
\usepackage{appendix}
\usepackage{float}
\usepackage{textcomp}
\usepackage{xcolor}
\usepackage{graphicx}
\usepackage{longtable}
\usepackage{blindtext, rotating}
\usepackage{ulem} 
\usepackage{amsmath} 

\DeclareMathOperator{\wrdexp}{exp}

\newcommand{\nobs} [1]{137}

\published{in PASP, 2018 Dec }
\shorttitle{Laser lines in Boyajian's star}
\shortauthors{Lipman et al.}

\begin{document}

\title{The Breakthrough Listen Search for Intelligent Life: Searching Boyajian's Star for Laser Line Emission}


\author{David Lipman}
\affiliation{Astronomy Department, University of California, Berkeley, CA, USA}
\affiliation{Kehillah Jewish High School,
Palo Alto, CA, USA}
\affiliation{Princeton, NJ 08544,USA}

\author{Howard Isaacson}
\affiliation{Astronomy Department, University of California, Berkeley, CA, USA}

\author{Andrew P. V. Siemion}
\affiliation{Astronomy Department, University of California, Berkeley, CA, USA}
\affiliation{ASTRON, Netherlands Institute for Radio Astronomy, Dwingeloo, NL}
\affiliation{SETI Institute, Mountain View, California, USA}
\author{Matt Lebofsky}
\affiliation{Astronomy Department, University of California, Berkeley, CA, USA}
\affiliation{Centre for Astrophysics \& Super-computing, Swinburne University of Technology, Hawthorn, VIC 3122, Australia}

\author{Danny C. Price}
\affiliation{Astronomy Department, University of California, Berkeley, CA, USA}
\affiliation{Centre for Astrophysics \& Super-computing, Swinburne University of Technology, Hawthorn, VIC 3122, Australia}

\author{David MacMahon}
\affiliation{Astronomy Department, University of California, Berkeley, CA, USA}
\author{Steve Croft}
\affiliation{Astronomy Department, University of California, Berkeley, CA, USA}
\author{David DeBoer}
\affiliation{Astronomy Department, University of California, Berkeley, CA, USA}
\author{Jack Hickish}
\affiliation{Astronomy Department, University of California, Berkeley, CA, USA}
\author{Dan Werthimer}
\affiliation{Astronomy Department, University of California, Berkeley, CA, USA}
\author{Greg Hellbourg}
\affiliation{Astronomy Department, University of California, Berkeley, CA, USA}

\author{J. Emilio Enriquez}
\affiliation{Astronomy Department, University of California, Berkeley, CA, USA}
\affiliation{Department of Astrophysics/IMAPP, Radboud University, Nijmegen, The Netherlands}

\author{Nectaria Gizani}
\affiliation{Astronomy Department, University of California, Berkeley, CA, USA}
\affiliation{School of Science and Technology, Hellenic Open University, Patra, Greece}


\begin{abstract}
	Boyajian' s Star (KIC 8462852) has received significant attention due to its unusual periodic brightness fluctuations detected by the Kepler Spacecraft and subsequent ground based observations. Possible explanations for the dips in the photometric measurements include interstellar or circumstellar dust, and it has been speculated that an artificial megastructure could be responsible. We analyze 177 high-resolution spectra of Boyajian's Star in an effort to detect potential laser signals from extraterrestrial civilizations. The spectra were obtained by the Lick Observatory's Automated Planet Finder telescope as part of the Breakthrough Listen Project, and cover the wavelength range of visible light from 374 to 970 nm. We calculate that the APF would be capable of detecting lasers of power greater than approximately 24 MW at the distance of Boyajian's Star, d = 1470 ly. The top candidates from the analysis can all be explained as either cosmic ray hits, stellar emission lines or atmospheric air glow emission lines.

\end{abstract}


\pagebreak

\section{Introduction}
\label{Introduction}

Among the many discoveries made from data collected from NASA's Kepler spacecraft \citep{Borucki2010}, non-periodic flux variations at the few percent level garnered significant attention \citep{Boyajian2016}. Many explanations have been put forth including the presence of a collection of planetisimals orbiting the star \citep{Bodman2016}, or perhaps a swarm of comet-like interstellar objects \citep{Makarov2016}. Observations from the UV to the mid-infrared conducted by \cite{Meng2017}, suggest the dimming is caused by circumstellar material. Ground based spectrophotometry conducted by \cite{Deeg2018} describe the dimmings as caused by optically thin dust with particle sizes of the order of 0.0015 to 0.15 \micron{}. \cite{Boyajian2018} have compiled ground based observations showing several distinct dimming events of a few percent each.  We follow up the idea set forth in \cite{WrightJ2016} suggesting that an alien megastructure constructed by an advanced civilization could be responsible for the flux variations. 

Boyajian's Star (also called KIC 8462852 or Tabby's Star) is an F3V star approximately 1,470 light years from Earth \citep{Gaia2018}, in the constellation Cygnus. The star experiences unusual non-periodic dimming of up to 20\% over a period of several days, which was first discovered in 2015 \citep{Boyajian2016}.  The objective of this project is to analyze stellar spectral data collected by Lick Observatory's Automated Planet Finder (APF) telescope to search for potential extraterrestrial laser signals from a technical civilization associated with  Boyajian's Star.

Radio SETI and optical SETI are the two primary observational components of the Breakthrough Listen program. The search for laser lines utilizes APF data while the Green Bank Telescope and Parkes Telescope are used to conduct the radio survey. The Breakthrough Listen Project's optical search uses time at the Lick Observatory's Automated Planet Finder (APF), a 2.4m aperture telescope on Mount Hamilton, California, to search for optical emission lines from a pre-determined list of nearby stars across the main sequence. Additionally, some of the time is reserved for targets of opportunity and exotica \citep{Isaacson2017} such as Boyajian's star.

This paper presents the analysis of 177 spectra of Boyajian's Star recorded between 2015 November 1 and 2018 May 3. These spectra will be valuable in the search, still in progress, for periodic variations in spectral features. Spectroscopic optical SETI programs take advantage of the prospective concentration of laser energy flux into a narrow beam and the tiny range of wavelengths to increase the ratio of laser flux to stellar flux \cite{Tellis2015,Tellis2017}, making them a potentially effective way to communicate across interplanetary and interstellar distances. 

In the search for laser lines we have developed an algorithm to perform pixel-by-pixel analysis of each spectrum to identify spatially unresolved emission lines that meet the criteria for an artificial laser signal. Those conditions include a full width half maximum (FWHM) close to the telescope's point spread function (PSF) and a wavelength that does not correspond to known terrestrial atmospheric air glow lines or naturally occurring stellar emission features. A signal injection-recovery process was used to measure our ability to detect signals. False positives resulting from cosmic ray events were eliminated through a secondary multi-step analysis process. We discuss our best candidates and provide non-ETI explanations for all of them.

\section{Observations}
\subsection{APF Spectra}

\label{APF Data section}

\begin{figure}[H]
\epsscale{0.70}
\plotone{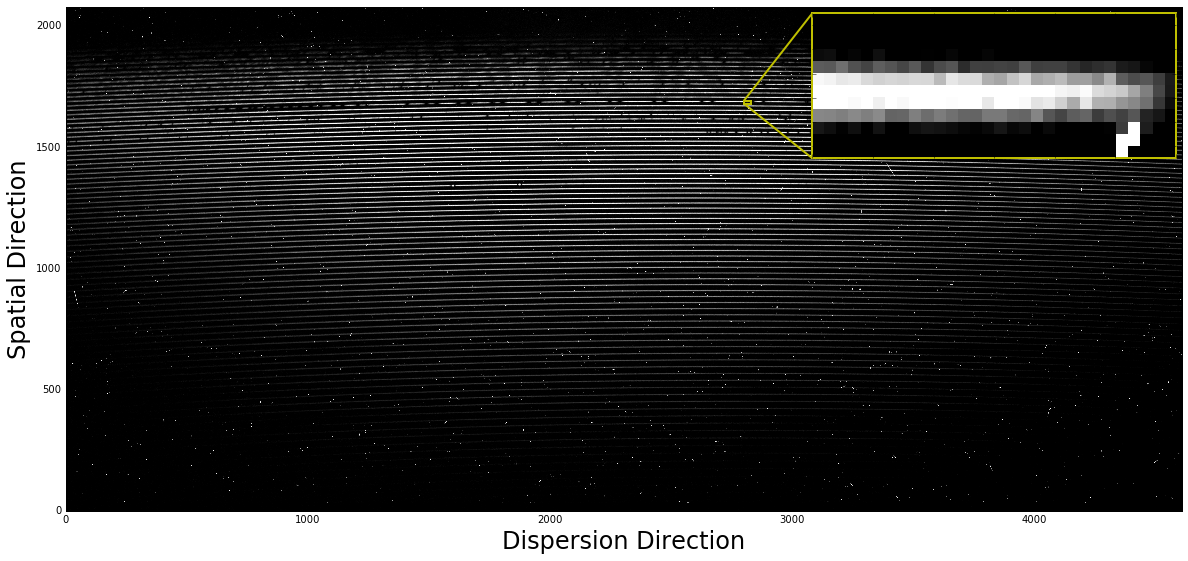}
\caption{A raw 2D echelle spectrum of Boyajian's Star, 4,608 pixels in the dispersion direction and 2,080 pixels in the spatial direction, containing 79 spectral orders. The zoomed image in the upper-right corner is a segment one of these spectral orders, which  shows a cosmic ray in the bottom right corner.}
\label{2D Spectrum}
\end{figure}

The Automated Planet Finder's sole instrument is the Levy Spectrometer \citep{Radovan2014,Vogt2014}. This high resolution, cross-dispersed echelle spectrometer creates a 2D spectrum containing 79 spectral orders, each spanning 4,608 pixels in the dispersion direction (Figure~\ref{2D Spectrum}). 
The APF spectra have a fixed format and wavelength range from $374 - 970$\,nm with a spectral resolution of approximately 100,000. The Levy Spectrometer provides a range of fixed-slit apertures on a high precision linear stage at the Nasmyth focus. Decker apertures range from angular sizes of $1\arcsec \times 12\arcsec$ to $8\arcsec \times 8\arcsec$. The two deckers in use for this project have dimensions of $1\arcsec \times 3\arcsec$ and $1\arcsec \times 8\arcsec$. Exposure times of either ten minutes or thirty minutes result in signal to noise ratios between twenty and thirty-five.

\begin{figure}[H]
\epsscale{0.70}
\plotone{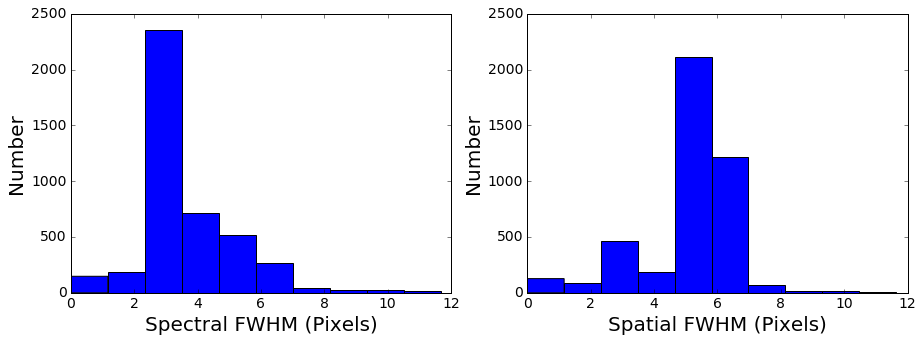}
\caption{A Thorium Argon spectrum taken with the $1\arcsec \times 3\arcsec$  decker was searched for peaks to calculate mean width of the APF instrumental point spread function. Left: A histogram of the spectral FWHM values. Right: A histogram of the spatial FWHM values. We refer to the dispersion direction as the X-direction and the spatial direction as the Y-direction in subsequent sections.}
\label{ThAr W Decker}
\end{figure}

\begin{figure}[H]
\epsscale{0.7}
\plotone{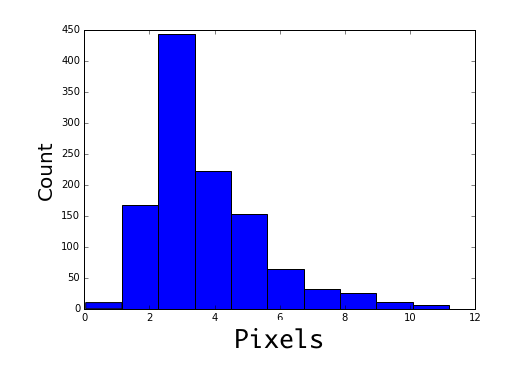}
\caption{Spatial FWHM of peaks detected in a Thorium-Argon spectrum captured using a pinhole decker. The median spatial FWHM is 3.23 pixels.}
\label{ThAr Pinhole}
\end{figure}

The instrumental point spread function (PSF) can be measured by analyzing pinhole and science decker thorium-argon spectra. Using 4,298 thorium-argon emission lines from a spectrum recorded with the $1\arcsec \times 3\arcsec$ science decker, we measure the median FWHM  equal to 3.26 pixels in the spectral dimension and 5.69 pixels in the spatial dimension (Figure~\ref{ThAr W Decker}). The pinhole decker results in FWHM values of 2.24 and 2.20 pixels in the wavelength and spatial directions, respectively (Figure~\ref{ThAr Pinhole}). On-sky conditions and use of the science decker create a larger point spread function, but the pinhole decker provides the lower limit on the PSF.  The use of the science decker and the local seeing conditions at the time of observation can cause the PSF to be larger, but not smaller than the thorium-argon lamp. Monochromatic extraterrestrial laser signals could also appear larger than the PSF due to the local seeing conditions, but only in the spatial direction.

\subsection{Radio Observations}
Breakthrough Listen and collaborators have acquired twelve hours of observations of Boyajian's star at the Green Bank Telescope in West Virginia. Taken using five different receivers covering the frequency range from 1.1 GHz to 12 GHz and 18-26 GHz these observations complement the optical spectra acquired with the APF. However the analysis of the radio data are beyond the scope of this paper.

\begin{figure}[H]
\epsscale{0.50}
\plotone{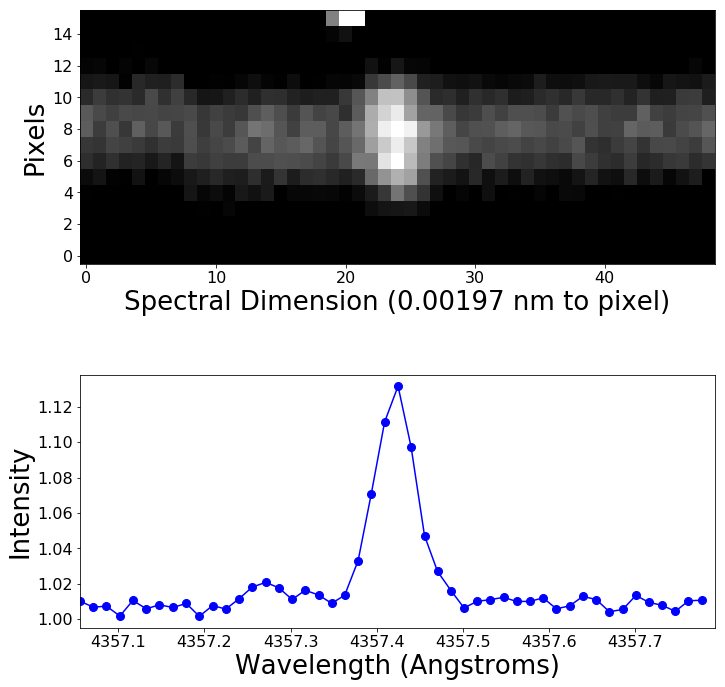}
\caption{APF Data: Top) A $50 \times 20$ pixel segment of a 2D echelle spectrum recorded by the Levy Spectrometer. Each column is summed to calculate the intensity for each wavelength value in the spectrum. Bottom) The reduced 1D spectrum.}
\label{APF Data}
\end{figure}

\section{Analysis}
\subsection{Spectrum Normalization}
\label{Spectrum Normalization}

Due to the optics of the APF and echelle spectrometers in general, each order of the 1D spectrum is brighter towards the center of the CCD than at the edges. This feature, the blaze function, can be removed to provide a spectrum with similar values across the continuum across all of the orders. To remove the blaze function, we divide each order into 100 pixel segments and calculate the 85th percentile value for each segment. We then fit a single third order polynomial to these values in each spectral order, and divide each point in the original reduced spectrum by the polynomial to generate a deblazed, normalized spectrum prior to analysis.

\subsection{Cosmic Rays}
\label{Cosmic Rays}

\begin{figure}[H]
\epsscale{0.50}
\plotone{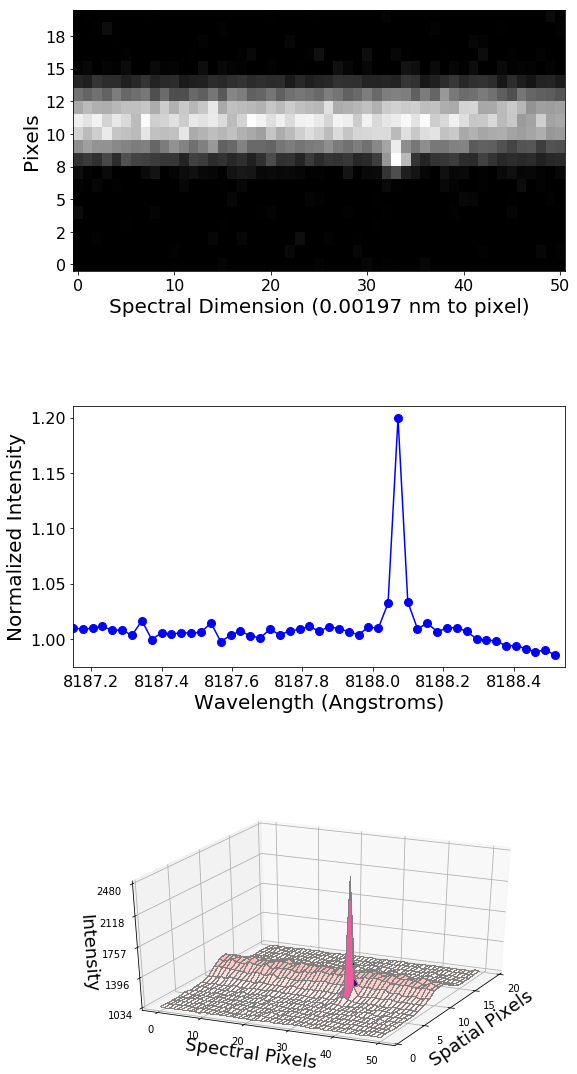}
\caption{Cosmic Ray Example - Top) A 2D display of one partial order showing a cosmic ray as the sharp spike in intensity at approximately pixel 32. Middle) A 1D plot of the vertically summed order with the cosmic ray and night sky emission feature. Bottom) A 3D model fit of the 2D display.}
\label{Cosmic Ray Ex}
\end{figure}

Cosmic rays are highly energetic particles (100 MeV to 100 GeV), typically from extra-galactic origin that continuously bombard the earth \citep{Funk2013}. Cosmic rays do not follow the light path through the telescope and spectrometer, and thus are not dispersed by the optical surfaces. Instead they directly impact the CCD detector (most commonly as highly ionizing muons created in cosmic-ray-induced atmospheric air showers). In Figure~\ref{Cosmic Ray Ex}, cosmic rays appear as high intensity pixels in the 2D spectrum with a profile narrower than the point spread function. As we search each spectrum for narrow-band peaks, cosmic rays are identified as a primary source of false positives, as discussed in Section~\ref{identify_fp}.

In order to detect potential signals of artificial origin, we developed an algorithm to search each of the 177 stellar spectra recorded by the APF for candidate signals. The algorithm contains three core components: Identification of peaks throughout the spectrum above a local mean flux, Gaussian fitting of every identified peak, and visual inspection of each Gaussian peak. These factors combine to form the comprehensive algorithm used in our search for laser lines from Boyajian's Star.

\subsection{Signal Detection Algorithm}
\label{Signal Detection Algorithm}

\subsubsection{Identify Local Peaks}
\label{Identify Local Peaks}

\begin{figure}[H]
\epsscale{0.50}
\plotone{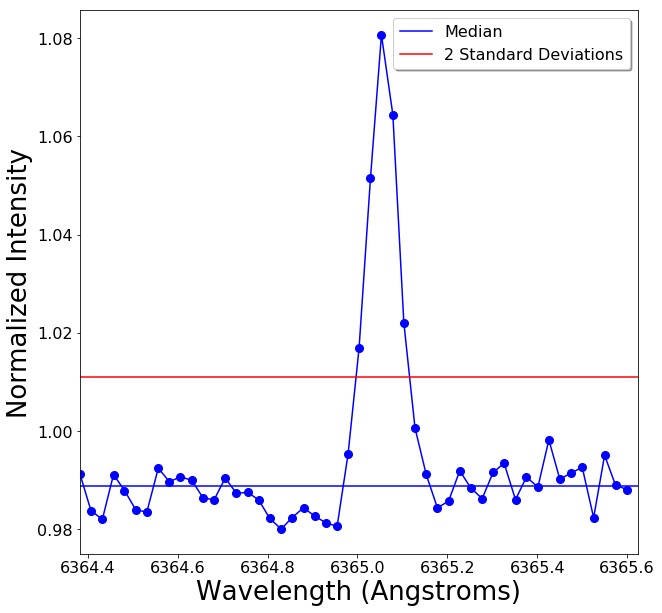}
\caption{Candidate signal rising above the 2-sigma threshold used to identify candidate signals.}
\label{Local Peak} 
\end{figure}

Each of the spectrum's 79 orders are analyzed in 100 pixel segments to identify and clip values that are at least 2 standard deviations above the local median. The standard deviation of the resulting dataset is calculated as the value of the local noise in the spectrum's background continuum. The local median and standard deviation are calculated using segments spanning out 50 pixels from the signal of interest in both directions. Figure~\ref{Local Peak} shows an example of a signal at least two standard deviations above its local median.

\begin{figure}[H]
\epsscale{1}
\plotone{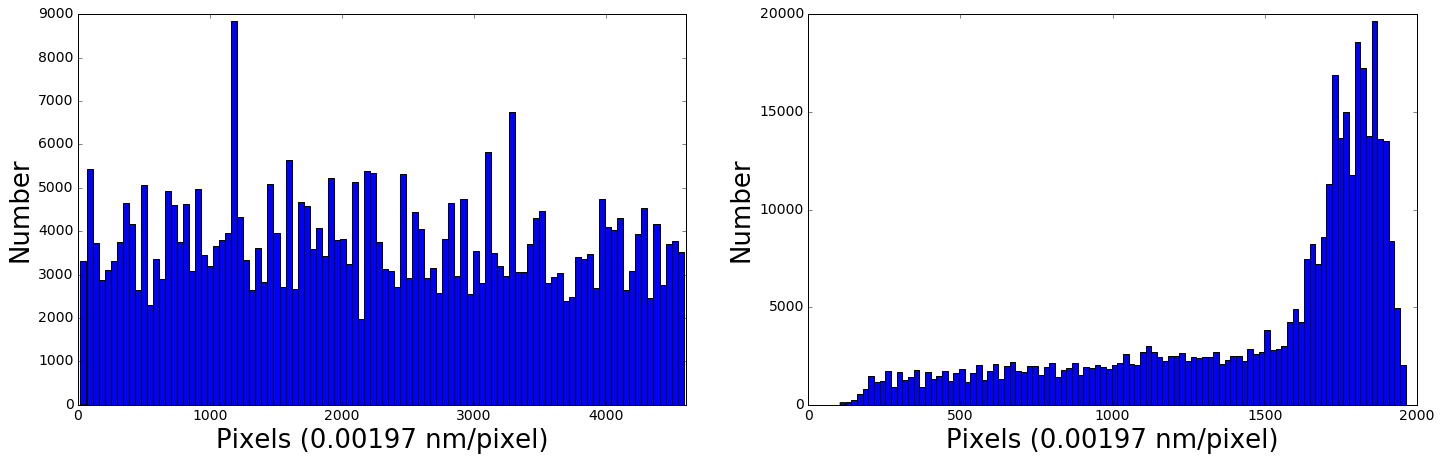}
\caption{Pixel number in the dispersion direction (left) and spatial direction (right) of all peaks identified by the algorithm prior to Gaussian fitting. 152,370 peaks were detected across the 177 spectra.}
\label{X_Y_Pos}
\end{figure}

\begin{figure}[H]
\epsscale{1}
\plotone{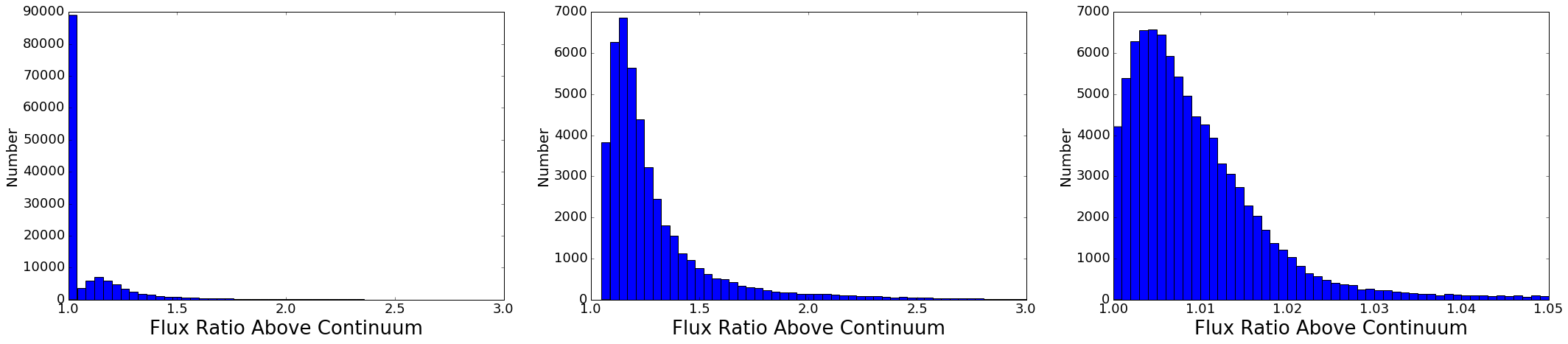}
\caption{Intensities of all 152,370 detected peaks measured in flux ratio above the continuum. From left to right, the domain of each of the histograms is: (1,3), (1.05,3), (1,1.05).}
\label{Intensities}
\end{figure}

Figure~\ref{X_Y_Pos} shows the distribution of 152,370 peaks detected across the 177 analyzed spectra. The peaks were evenly distributed in the X-dimension of the 2D spectrum. They are concentrated between 1,500 and 2,000 pixels in the Y-dimension, which corresponds to the peak in the black-body spectrum for a star of this spectral type. The intensities of these peaks were highly concentrated in between 1 and 1.05 times the local continuum (Figure~\ref{Intensities}). Signals with small amplitudes such as these are found in regions with few spectral absorption features allowing the algorithm to capture many insignificant signals. None of our top candidates are in this range of intensities between 1.0 to 1.05 times the continuum.

A 2D Gaussian function is fit to each local peak to calculate the full width half maximum (FWHM) value in the spatial and spectral directions. The continuum level flux is subtracted before the FWHM calculation is performed. Those peaks with spectral FWHM of an upper-bound of 3.53 pixels and lower-bound of 2.36 pixels, and spatial FWHM of an upper-bound of 5.77 pixels and lower-bound of 3.23 pixels (PSF approximate upper-bounds are 3.5 and 5.8 pixels respectively) are kept for further analysis. Any candidate with a spatial FWHM less than 3.23 pixels was discarded as a false positive, since any profile that is narrower than that of the pinhole decker does not share the PSF properties of the spectrometer.

\subsubsection{Identify False Positives}
\label{identify_fp}

\begin{figure}[H]
\epsscale{1}
\plotone{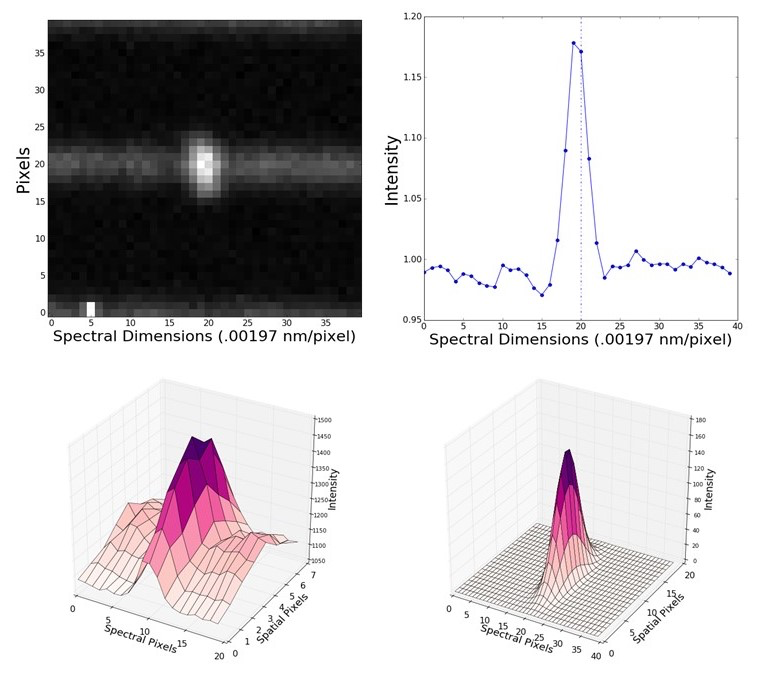}
\caption{Atmospheric Airglow Line. Left) 2D airglow line. 2nd from left) 1D signal visualization, 3rd from left) 3D visualization of the data prior to Gaussian fitting. Right)   3D visualization of the best-fit Gaussian model, including spatial and spectral FWHM values.}
\label{Airglow}

\end{figure}

False positives resulting from cosmic ray events and other false alarms are eliminated through a multi-step analysis process: 

\begin{enumerate}

\item Each signal's wavelength is compared against the European Southern Observatory (ESO) database of known atmospheric airglow lines\footnote{http://www.eso.org/sci/facilities/paranal/decommissioned/isaac/tools/lib.html}.  Atmospheric airglow lines are resolvable spatially, as opposed to stars which are point sources. Figure~\ref{Airglow} shows 1D, 2D, and 3D visualizations of an atmospheric airglow line, as well as the signal's Gaussian fit. In the 2D visualization, the airglow line is broader than the stellar profile in the spatial dimension. The algorithm avoids most airglow lines, as they are spatially broader than the PSF. A spatial FWHM threshold of an upper-bound of 5.770 pixels is used to dismiss potential airglow lines.

\item As discussed in Section \ref{Cosmic Rays}, cosmic rays cause apparent features in the spectrum that are narrower than the PSF. A minimum FWHM threshold of 2.36 pixels in the spectral dimension and 3.23 pixels in the spatial dimension is used to exclude false positives from cosmic rays. The 3D representation of each candidate signal is further inspected visually to identify false positives resulting from cosmic rays or other outliers.

\item Candidate signals that appear in multiple spectra at the same wavelength are evaluated and their absolute positions on the CCD are compared across observations. Signals that fall onto the same pixels when the stellar lines are shifted due to varying barycentric corrections are definitively atmospheric, as extraterrestrial laser signals would be Doppler shifted by the barycentric motion of the Earth and would always appear at the same wavelength relative to other stellar absorption lines. 
\end{enumerate}

\subsection{Signal Injection and Recovery}
\label{Signal Injection}

\begin{figure}[H]
\epsscale{1}
\plotone{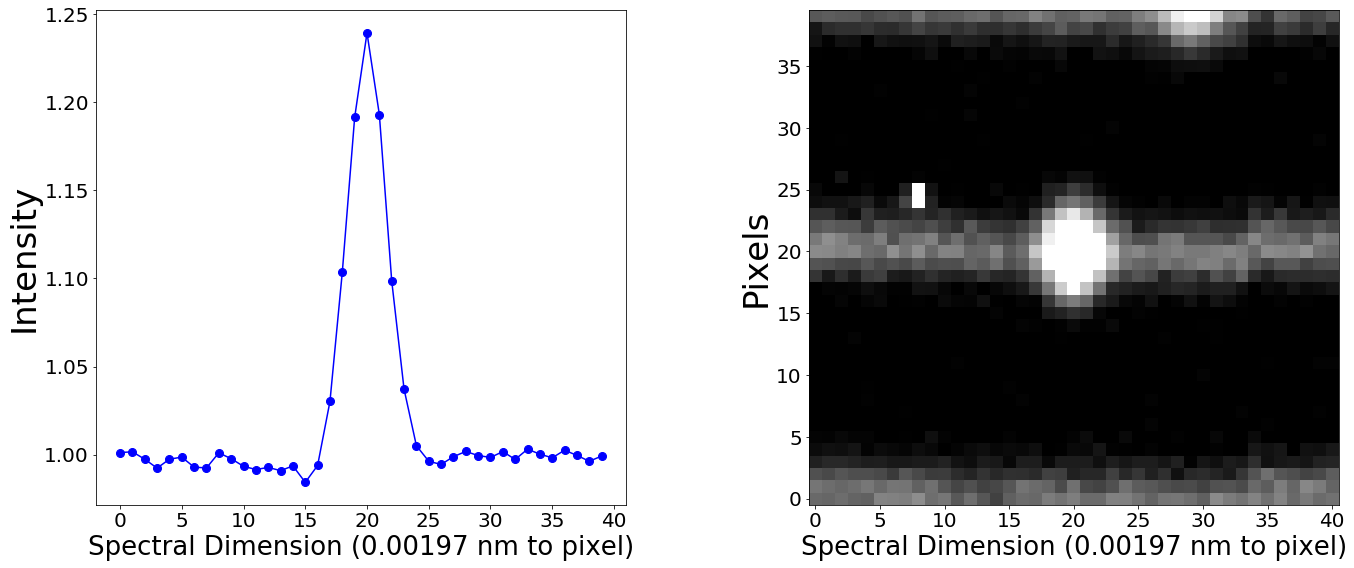}
\caption{Injected Signal - The figure on the left is a 1D visualization of an injected signal. The figure on the right shows an expanded view of the injected signal in the 2D spectrum.}
\label{Injected Signal}
\end{figure}

In order to characterize the laser detection algorithm, we injected 20,000 simulated laser signals into 20 raw 2D spectra. Signals are simulated using 2D Gaussian curves with parameters expected for a laser signal, and were injected at random locations along the orders within the spectra, across a range of signal-to-noise (SNR) ratios from 3 to 500 (Figure~\ref{Injected Signal}). The algorithm detects 98.29\% of the injected signals with SNR greater than 20, giving high confidence that the detection algorithm is capable of accurate identification of extraterrestrial lasers above this SNR value. 

\begin{figure}[H]
\epsscale{1}
\plotone{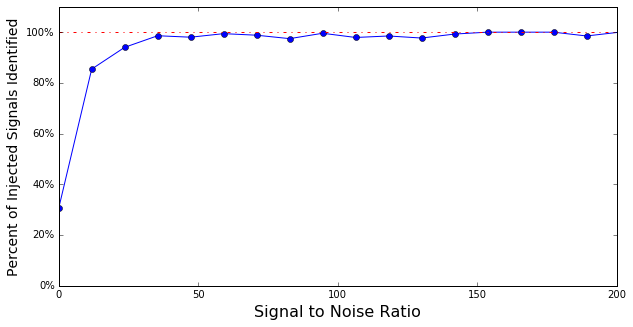}
\caption{The recovery rate of injected signals at varying SNRs is 98.29\%  at SNR = 20 and 99.9\% above SNR = 52.}
\label{Recovery}\begin{flushleft}
\end{flushleft}
\end{figure}

\subsection{Candidate Extraterrestrial Laser Signals}
\label{Candidate Extraterrestrial Laser Signals}

\begin{figure}[H]
\epsscale{1}
\plotone{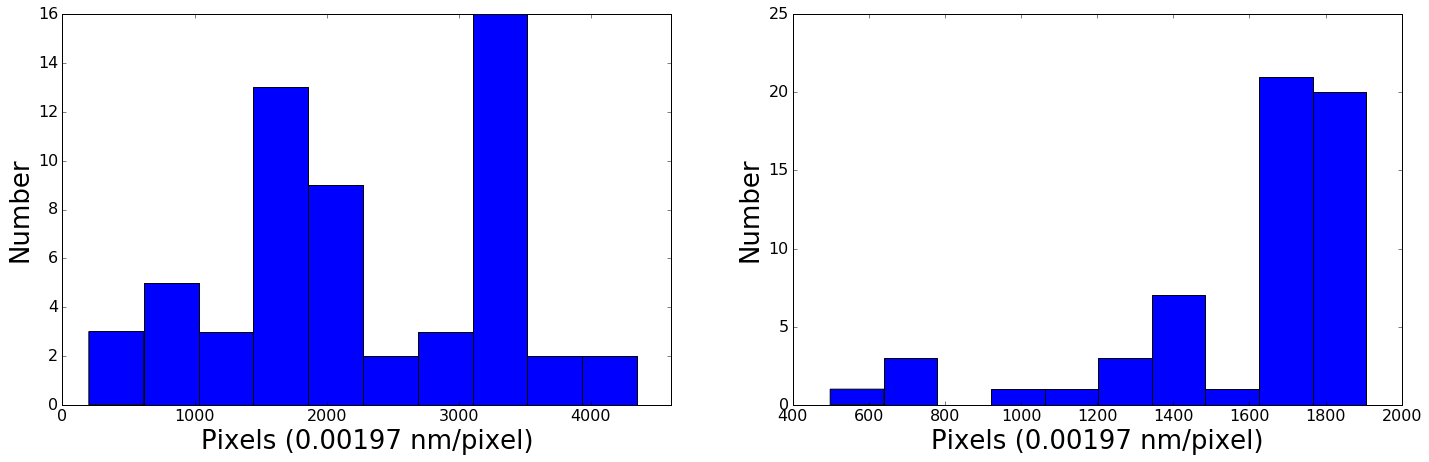}
\caption{Dispersion (left) and spatial (right) position (as discussed in Figure \ref{2D Spectrum}) of each of the 40 candidate signals identified by the algorithm. The X positions of candidate signals are most highly concentrated in two regions; between a pixel position 1,000 to 2,000, and between pixel positions 3,000 to 3,500. These regions of concentration contrast with the random spread of the X positions of all hits.}
\label{Candidate_Pos}
\end{figure}

\begin{figure}[H]
\epsscale{1}
\plotone{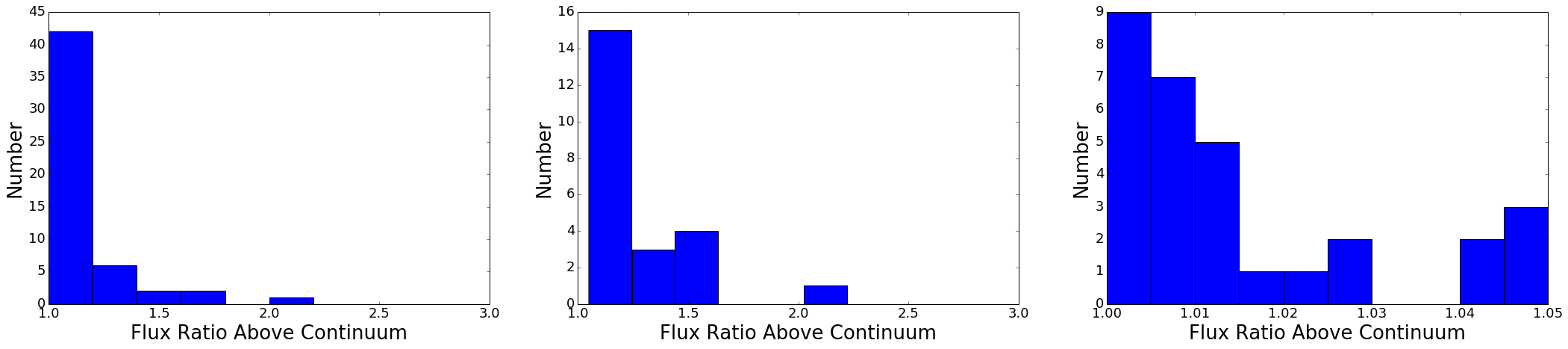}
\caption{Intensities of each of the 40 candidate signals identified by the algorithm. From left to right, the domain of the histograms are: (1,3), (1.05,3), (1,1.05). The majority of the intensities of the candidates are highly concentrated between 1 and 1.05 times the local continuum.}
\label{Candidate_Intensity}
\end{figure}

The laser search algorithm detects 58 candidate signals in the 177 Boyajian's Star spectra, of which 53 were eliminated as being false positives. Of these 53 false positives, 11 were discarded as being atmospheric airglow lines, 15 were discarded as being cosmic rays, and 27 were discarded as being random fluctuations in the stellar continuum. Figure \ref{Candidate_Pos} shows the X and Y position of each of the candidates on the 2D spectrum, while Figure \ref{Candidate_Intensity} shows the intensities of each of the candidates.

\begin{deluxetable*}{llllllllll}
\tablewidth{0pt}
\tabletypesize{\scriptsize}
\tablecaption{Properties of Top 5 Candidate Signals \label{tab:stars_table} }
\tablehead{
\colhead{Date} &
\colhead{Observation} &
\colhead{SNR} &
\colhead{Wavelength}  & 
\colhead{Spectral}  &
\colhead{Spatial} &
\colhead{Order} &
\colhead{Pixel Number} &
\colhead{Amplitude}\\ 
& 
\colhead{Name} &
& 
\colhead{(nm)} &
\colhead{FWHM}  & 
\colhead{FWHM} & 
\colhead{Number} &
\colhead{Spatial} &
\colhead{Spatial}
}
\startdata
2016-04-04 & ucb-aoz236 & 4.48 & 850.816 & 3.32 & 5.76 & 69& 3218 & 1.07\\
2016-06-06 & ucb-aqd243 & 5.63 & 589.076 & 3.29 & 5.72 &45&1597&2.11\\
2016-06-06 & ucb-aqd243 & 11.87 & 636.506 & 3.48 & 5.48 &51&1191&1.12\\
2016-08-14 & ucb-art205 & 13.01 & 636.505 & 3.22 & 5.68 &51&1191&1.08\\
2016-10-23 & ucb-ata189 & 3.76 & 818.645 & 3.46 & 5.72 &67&2350&1.60\\
\enddata
\end{deluxetable*}

The five remaining candidates were identified at 636.5 nm, 818.6 nm, 589.1 nm, and 850.8 nm, as shown in Table \ref{tab:stars_table}. The 636.5 nm  signal was identified in two spectra, while the other candidates were only identified in a single spectrum. Each of the candidates has a spectral FWHM of between 2.35 and 3.53 pixels and a spatial FWHM between 3.23 and 5.77 pixels. The candidates are furthermore not wavelengths listed in the ESO's database of known atmospheric emission lines.

All 177 spectra were analyzed for occurrences of peaks at the candidate wavelengths. Peaks at these wavelengths are common in many of the spectra across varying barycentric corrections. If the candidate signals were of extraterrestrial origin they would be Doppler shifted as a function of barycentric correction, and we would expect to see a sharp drop-off in SNR at the specific candidate wavelength. As peaks were identified at each wavelength across varying barycentric correction, we must conclude that these candidates are terrestrial in nature.

\begin{figure}[H]
\epsscale{0.7}
\plotone{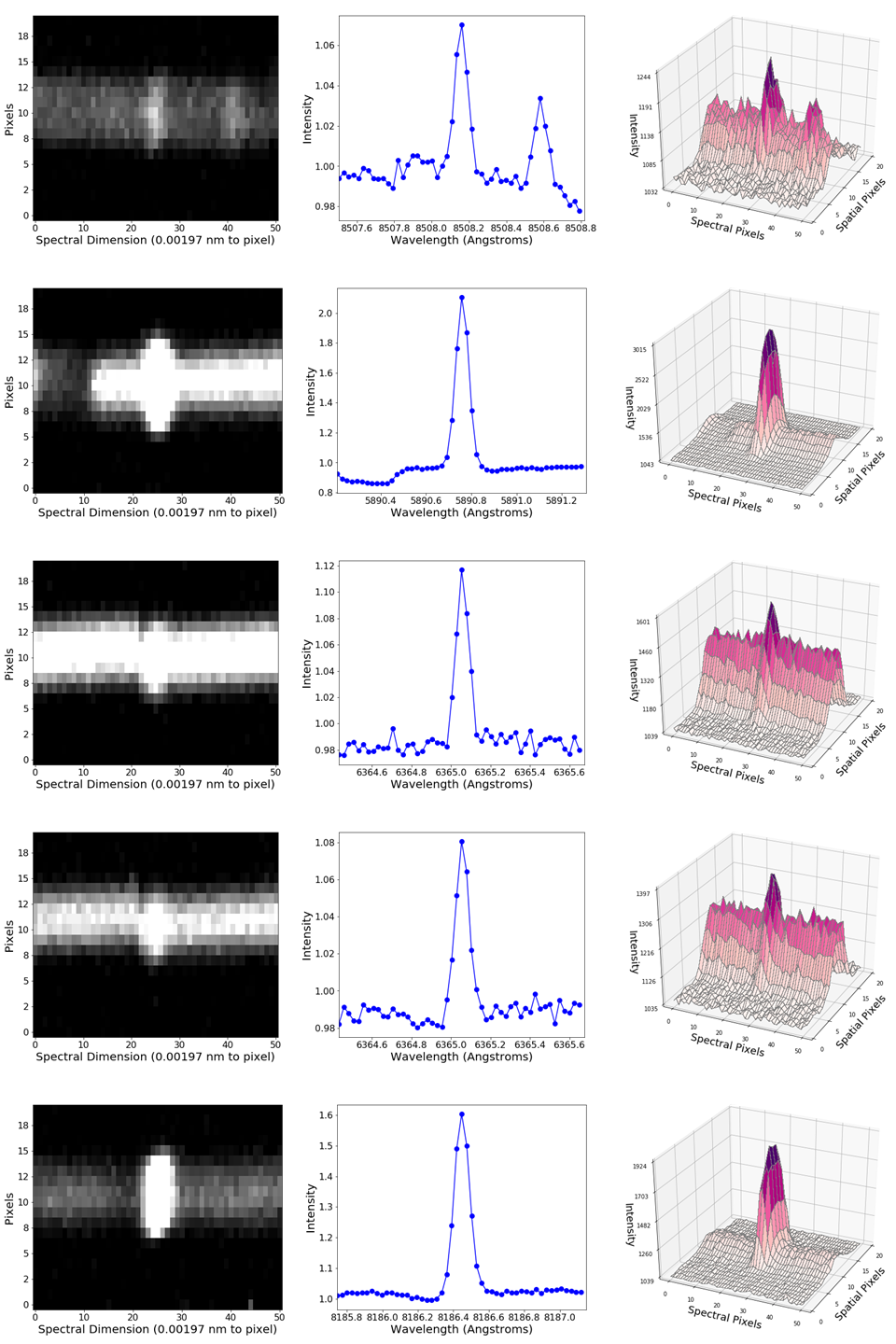}
\caption{Candidate Signals: 2D, 1D, and 3D representations of each of the 5  candidate signals. From bottom to top, these signals were detected at 850.816 nm, 589.076 nm, 636.506 nm, 636.505 nm, and 818.645 nm respectively. We attribute each signal to atmospheric airglow.}
\label{Candidate Signals} 
\end{figure}

\section{Discussion}
\subsection{Laser Detectability at Interstellar Distances}
\label{Laser Detectability at Interstellar Distances}

Lasers within the range of current technology are detectable across interstellar distances. A high resolution spectrograph, such as the APF-Levy, will enable discrimination between lasers and signals of natural origin such as stellar emission lines. Careful selection of signal detection parameters (signal to noise ratio, peak shape, etc) may yield potential candidate signals for deeper analysis and study.

For the simplified case of a smooth stellar spectrum dominated by Poisson noise, detection of a laser requires that its power be greater than the underlying noise of the part of the star's continuum in which it appears. Since light from a laser spreads out due to diffraction, lasers become weaker as the inverse of the distanced squared. For a star of sun-like luminosity, with a laser transmitter the same aperture as the APF at a distance of 1,470 light-years, the range of detectable power is approximately 24\,MW. The derivation and details are explained in the appendix. 

\section{Conclusion}
In this paper we have analyzed 177 spectra of Boyajian's Star for extraterrestrial laser signals using the Lick Observatory's Automated Planet Finder. We calculated that the Levy Spectrometer can detect continuous laser emission of power greater than approximately 24\,MW. This power is within the capabilities of modern human technology. As lasers are a relatively recent development in human history, it is not inconceivable that extraterrestrial civilizations would have developed lasers with comparable or potentially significantly greater power.

Other than a laser beacon deliberately directed by an extraterrestrial civilization towards Earth, it is to be expected that extraterrestrial laser signals would be transient in nature. This is due to likely relative motion of an emitter with respect to Earth, and conceivably the likely use cases for high powered lasers (for example: communication, spaceship propulsion, military). To maximize the probability of detecting extraterrestrial laser signals, a variety of stellar types and the stars most nearby the sun should be observed and their spectra searched for laser lines.

\section{Acknowledgements}
\label{Acknowledgements}

We thank Gloria and Ken Levy for support of the Automated
Planet Finder Spectrometer. We are grateful to the Berkeley SETI Research Center (BSRC) for their support of this research.  We thank the staff and students of Breakthrough Listen for their support of SETI data collection and making this data publicly available at (\href{http://breakthroughinitiatives.org/opendatasearch}{breakthroughinitiatives.org/opendatasearch}).  Breakthrough Listen is funded by the Breakthrough Initiatives (\href{http://breakthroughinitiatives.org}{breakthroughinitiatives.org}).  Research at Lick Observatory is partially supported by a generous gift from Google. 

\appendixpage

Current technology can produce lasers which are detectable across interstellar distances. A high resolution spectrograph, such as the Levy, will enable discrimination between lasers and signals of natural origin such as stellar emission lines. Careful selection of signal detection parameters (signal to noise ratio, peak shape, etc) may yield potential candidate signals for deeper analysis and study. The ability to detect a hypothetical laser signal depends on the properties of the spectrograph, the telescope, the emitter, and the background, e.g. the star. Here we present two methods for determining the detection limit of our algorithm. 

Method 1:
In order to detect a laser line above the background continuum from  the star, the number of photons from the laser, $N_{l}$, must exceed the continuum of the star in the specified wavelength bin (defined by the width of the point spread function (PSF)).  Using Poisson statistics, a $10\sigma$ detection requires that the number of photons received from the transmitting laser be $10~\sqrt[]{N_s}$, where $N_s$ is the number of photons received from the transmitter's host star. 

First the number of photons expected from the laser is computed, assuming the emitter to be monochromatic. Then we compute the number of photons that we receive from the star in a specific wavelength bin. 

The number of photons from the laser received by the telescope and spectrograph is given by: 

\begin{equation}
N_{l} = \frac{F_{l} ~ A_{d} ~ t_{\wrdexp}} {E} \epsilon
\end{equation}

where $F_l$ is the flux from the laser, $A_{d}$ is the area of the receiving telescope, $t_{\wrdexp}$ is the exposure time in seconds, $E$ is the energy per photon, and $\epsilon$ is the efficiency of the Levy. 
The flux from the laser is defined as the luminosity of the laser $L_l$ divided by the area over which the laser emits, $A_l$, which is equal to the . The opening angle of the laser is $\Theta = \frac{1.22 \lambda}{d_t}$, where $d_t$ is the diameter of the transmitting telescope. We adopt a value for the emitting telescope, $d_t$, to be equal to the diameter of the APF. The area over which the laser emits is $A_l = \pi r^2 = \pi (D\Theta/2)^2$

\begin{equation}  
F_{l} = \frac{L_{l} }{A_{l}} 
= \frac{4L_{l}}{ \pi  (D\frac{1.22\lambda}{ d_{t}})^{2} }
\end{equation}

The definition of the flux of laser can be inserted into Equation 1. For a laser of luminosity $L_{l}$ operating at wavelength, $\lambda = 500$ nm, transmitted by an aperture with diameter $d_{t}=2.4 $ m over a distance $D=1470$ ly, the number of photons emitted by the laser is given by:

\begin{equation}
N_{l} 
= \frac{4L_{l} d_{t}^2}{1.22^{2}\pi hc  \lambda  D^{2}} ~ A_d ~ t_{\wrdexp} ~\epsilon
\end{equation}

If we assume for simplicity the approximate spectral energy distribution of a sun-like star as constant over a wavelength range, $\Delta \lambda$ = 400--900~nm, and $\delta \lambda$  = 3.26 pixels times 0.02 nm per pixel as the  wavelength range contained within a point spread function (PSF), the number of photons received from the star, $N_{s}$, at wavelength $\lambda$,  is given by

\begin{equation}
N_{s}  = \frac{L_s}{4\pi D^{2}} ~ \frac{\delta \lambda}{\Delta \lambda} ~ \frac{\lambda}{hc}  ~ A_{d} ~ t_{\wrdexp} ~\epsilon
\end{equation}

Returning to our detection requirement of $N_{l} > 10 \sqrt{N_{s}}$, and inserting our equations for $N_{l}$ and $N_{s}$ solving for the luminosity of the laser and we find:

\begin{equation}
L_{l} 
> 
\frac{1.86 ~ D}{ d_t^2   } \left( L_{s} \frac{ \delta \lambda}{\Delta \lambda} \frac{hc~\pi~ \lambda^3 } {A_{d}~t_{\wrdexp}~\epsilon} \right)^{1/2}
\end{equation}

For our observations with the APF, $t_{\wrdexp}=600$s, and we assume an APF-sized transmitter and receiver. We have approximated the luminosity of the star with the luminosity of the sun, $L_s = 3.828 \times 10^{26} $ W. $A_d$, the area of the receiving telescope is $\pi (1.2 m)^2=4.53~ m^2$ and $\lambda = 500$ nm. As such, for a transmission from a distance of $1470$ ly, and an efficiency of 5\%, our detection threshold is 24 MW. 

The primary reason that we can detect a laser over the flux of a star is because of the focused nature of the laser compared to the isotropic emission of the star. The second reason is the very narrow wavelength range over which the laser emits compared to the broad wavelengths range over which a sun-like star emits its radiation. 

Method 2: We present a limit on detectable lasers that is based upon the representative observation of Boyajian's star and the number of photons collected in a single PSF.

Consider one of our observations, ten minutes in duration, resulting in roughly 500 photons per pixel, and $3.26 \times 500$ photons per pixel equal to 1650 photons in a PSF. 3.26 is the FWHM of the PSF in the spectral direction, the number of pixels over which any point source would be spread. The dominant source of noise, Poisson noise, is the square root of the number of photons. Our 10-sigma threshold becomes $N_l=10 \times \sqrt{1650} = 406$. 

At this point, we have an empirical measure of the number of photons required to be emitted by the laser.
Using the same instrumental details above in Equation 3 to solve for the luminosity of a detectable laser $(L_l)$, we find a threshold of 11.7 MW.  This agrees to within a factor of two with the result from Method 1.


\bibliographystyle{apj}
\bibliography{targets.bbl} 

\end{document}